\documentclass[12pt]{article}
\usepackage{amsmath}
\usepackage{enumerate}
\usepackage{amsfonts}
\begin{document}
\def\be{\begin{equation}}
\def\bea{\begin{eqnarray}}
\def\ee{\end{equation}}
\def\eea{\end{eqnarray}}
\def\d{\partial}
\def\eps{\varepsilon}
\def\la{\lambda}
\def\b{\bigskip}
\def\tgt{}  
\def\nn{\nonumber \\}
\def\t{\tilde}
\def\varkappa{\kappa}

\makeatletter
\def\blfootnote{\xdef\@thefnmark{}\@footnotetext}  
\makeatother

\begin{center}
{\LARGE Supergravity background\\ ~\\ of the $\la$--deformed AdS$_3\times$S$^3$ supercoset}
\\
\vspace{18mm}
{\bf   Yuri Chervonyi  and   Oleg Lunin}
\vspace{14mm}

Department of Physics,\\ University at Albany (SUNY),\\ Albany, NY 12222, USA\\ 

\vskip 10 mm

\blfootnote{ichervonyi@albany.edu,~olunin@albany.edu}

\end{center}

\begin{abstract}

We construct the solution of type IIB supergravity describing the integrable $\la$--deformation of the AdS$_3\times$S$^3$ supercoset. While the geometry corresponding to the deformation of the bosonic coset has been found in the past, our background is more natural for studying superstrings, and several interesting features distinguish our solution from its bosonic counterpart. We also report progress towards constructing the $\la$--deformation of the AdS$_5\times$S$^5$ supercoset.

\b

\end{abstract}

\newpage

\tableofcontents

\section{Introduction}
 
Integrability is a remarkable property, which has led to a very impressive progress in understanding of string theory over the last two decades (see \cite{IntReview} for review). While initially integrability was discovered for isolated models, such as strings on AdS$_p\times$S$^q$ \cite{IntCases}, later larger classes of integrable backgrounds have been constructed by introducing deformations parameterized by continuous variables. The first example of such family, known as beta deformation \cite{betaFT}, has been found long time ago \cite{beta}, but recently two new powerful tools for constructing 
integrable string theories have emerged. One of them originated from studies of the Yang--Baxter sigma models \cite{BosonicYangBaxter,Qdeform,Yoshida}, and it culminated in construction of new integrable string theories, which became known as $\eta$--deformations \cite{Delduc1,OtherEta,LRT,MoreEta,MoreEta2}. The second approach originated from the desire to relate two classes of solvable sigma models, the Wess--Zumino--Witten \cite{WZW} and the Principal Chiral \cite{PCM} models, and it culminated in the discovery of a one--parameter family of integrable conformal field theories, which has WZW and PCM as its endpoints \cite{SftsGr,HMS,ST}\footnote{See \cite{PreST} for earlier work in this direction.}. This connection becomes especially interesting when the PCM point represents a string theory on AdS$_p\times$S$^q$ space, and the corresponding families, which became known as $\la$--deformations, have been subjects of recent investigations \cite{SfetsosAdS5,HT,Holl2,BTW}. A close connection between the $\eta$ and $\la$ deformations has been demonstrated in \cite{HT}. In this article we study the $\la$--deformation for AdS$_3\times$S$^3$ and AdS$_5\times$S$^5$.

While the metrics for the $\la$--deformation of AdS$_p\times$S$^q$ have been constructed in \cite{ST,SfetsosAdS5}, the issue of the fluxes supporting these geometries has not been fully resolved. Although the metric for the deformation can be uniquely constructed starting from the corresponding coset,  there are two distinct prescriptions for the dilaton: one is based on a bosonic coset \cite{ST}, and the other one uses its supersymmetric version \cite{HMS}. In the first case the deformations for all AdS$_p\times$S$^q$ have been constructed in a series of papers \cite{ST,SfetsosAdS5}, while in the second case, which is more natural for describing superstrings, only the result for AdS$_2\times$S$^2$ is known \cite{BTW}. In this article we construct the geometry describing the $\la$--deformed AdS$_3\times$S$^3$ supercoset and report progress towards finding the deformed AdS$_5\times$S$^5$ solution. 

This paper has the following organization. In section \ref{SecReview} we review the procedure for constructing the $\la$--deformation, which will be used in the rest of the paper. In section \ref{SecAdS3} we use this procedure to construct the metric and the dilaton for the deformed AdS$_3\times$S$^3$, but unfortunately construction of Ramond--Ramond fluxes requires a separate analysis. In section \ref{SecAdS3RR} we determine these fluxes by solving supergravity equations, and in sections \ref{SecAdS3SpecCases}--\ref{SecAltParamAdS3} we find some interesting connections between the new background and solutions which exist in the literature. Section \ref{SectAdS5} reports progress towards constructing the $\la$--deformation for super--coset describing strings on AdS$_5\times$S$^5$. Specifically, we determine the metric and the dilaton, but unfortunately we were not able to compute the Ramond--Ramond fluxes. The $\la$--deformation of AdS$_2\times$S$^2$ constructed in \cite{BTW} is reviewed in Appendix \ref{AppAdS2}, and its comparison with higher dimensional cases is performed throughout the article.

\section{Brief review of the $\la$--deformation}
\label{SecReview}

\renewcommand{\theequation}{2.\arabic{equation}}
\setcounter{equation}{0}

We begin with reviewing the procedure for constructing the NS--NS fields for the $\la$--deformed cosets.  Such deformation belongs to a general class of two--dimensional integrable systems with equations of motion in the form
\bea\label{EOMcrnt}
&&\d_\mu I^\mu=0,\nn
&&\d_\mu I_\nu-\d_\nu I_\mu+[I_\mu, I_\nu]=0,
\eea
where currents $I_\mu$ take values in a semi--simple Lie algebra. Integrability of this system can be demonstrated by writing it as a zero--curvature condition for a linear problem\footnote{We denote that spectral parameter by $\Lambda$ instead of the conventional $\la$ to avoid confusion with a variable governing the deformation.}:
\bea
&&\mathcal{D}_\mu\Psi=0,\qquad \mathcal{D}_\mu(\Lambda)=\d_\mu+\frac{\Lambda^2}{\Lambda^2-1}I_\mu+\frac{\Lambda}{\Lambda^2-1}\epsilon_{\mu\rho}I^\rho,\nn
&&[\mathcal{D}_\mu(\Lambda),\mathcal{D}_\nu(\Lambda)]=0\,.\quad
\eea
Two well--known examples of the integrable systems described by equations (\ref{EOMcrnt}) are the  Principal Chiral Model (PCM) \cite{PCM} and the Wess-Zumino-Witten model \cite{WZW} for a group $G$:
\bea\label{PCM}
S_{PCM}(\tilde{g})&=&-\frac{\kappa^2}{\pi}\int\mbox{Tr}(\tilde{g}^{-1}\d_+\tilde{g} \tilde{g}^{-1} \d_-\tilde{g}),\quad I_\mu=\tilde{g}^{-1}\d_\mu \tilde{g},\\ 
\label{WZW}
S_{WZW}(g)&=&-\frac{k}{2\pi}\int\mbox{Tr}\left( g^{-1}\d_+g g^{-1}\d_- g \right)+\frac{ik}{6\pi}\int_B \mbox{Tr}(g^{-1}d g)^3,\quad I_\mu=g^{-1}\d_\mu g,
\eea
and the $\la$--deformation interpolates between these systems. This deformation utilizes two important symmetries of (\ref{PCM}) and (\ref{WZW}): the global $G_L\times G_R$ symmetry of the PCM and the  $G_{L,cur}\times G_{R,cur}$ symmetry of the current algebra of the WZW.

\bigskip
\noindent
{\bf $\la$--deformation for groups}

Let us review the construction introduced in \cite{SftsGr}, which allows one to interpolate between the systems (\ref{PCM}) and (\ref{WZW}) while preserving integrability. To find such $\la$ deformation, one adds the PCM and WZW models (\ref{PCM}), (\ref{WZW}) for the same group $G$ and gauges the $G_L\times G_{diag,cur}$ subgroup of global symmetries. This is accomplished by modifying the derivative in the PCM as 
\bea
\d_\pm\tilde{g}\to D_\pm\tilde{g}=\d_\pm\tilde{g}-A_\pm\tilde{g},
\eea
and by gauging the resulting WZW model. Integrating out the gauge fields $A_\pm$, one arrives at the final action \cite{SftsGr}\footnote{We follow the conventions of \cite{HT,BTW}, and the deformation parameter $\la$ used in \cite{SftsGr,ST} is equal to $\la^2_{our}$.}
\bea\label{ActDeform}
S(g)&=&S_{WZW}(g)+\frac{1}{\pi}\frac{k^2}{k+\kappa^2}\int J_+^a(1-\lambda^2 D)^{-1}_{ab}J^b_-,\quad
\lambda^2=\frac{k}{k+\kappa^2},\quad 0\le\lambda\le 1,\\
 D_{ab}&=&\mbox{Tr}(t_a g^{-1} t_b g),\quad J_\pm^a=-i\mbox{Tr}(t^a \d_\pm g g^{-1}):\quad J_+^a=R_\mu^a\d_+ X^\mu,\quad J_-^a=L_\mu^a\d_- X^\mu\,.\nonumber
\eea
Deformation (\ref{ActDeform}) interpolates between the PCM ($\lambda=1$) and the WZW model ($\lambda=0$) while preserving integrability \cite{SftsGr}.
 
 To extract the gravitational background describing the deformation, one rewrites (\ref{ActDeform}) as
\bea
S(g)=S_{WZW}(g)+\frac{k^2}{\pi}\int (R^T  M^{-1} L)_{\mu\nu} \d_+ X^\mu \d_- X^\nu,\qquad
M=(k+\kappa^2)(1-\lambda^2 D),
\eea
and compares the result with the action of the sigma model
\bea
S=\frac{1}{2}\int (G+B)_{\mu\nu}\d_+ X^\mu \d_- X^\nu\,.
\eea
This leads to the metric and to the Kalb-Ramond field:
\bea\label{GeneralMetric}
ds^2&=&\frac{k}{2\pi}L^T L+\frac{k^2}{2\pi}L^T(DM^{-1}+[M^{-1}]^TD^T)L\\
B&=&\frac{1}{1-\lambda^4}\left( B_0+\frac{\lambda^2}{2}L^T\left[(D^T-\lambda^2)^{-1}-(D-\lambda^2)^{-1} \right]\wedge L\right),\nonumber
\eea
where $B_0$ is a Kalb-Ramond field of an undeformed WZW model with the field strength
\bea
H_0=-\frac{1}{6}f_{abc}L^a\wedge L^b\wedge L^c.
\eea
Recalling the definition of $M$ and the relation $D^TD=1$, one can rewrite the metric in terms of convenient frames:
\bea
ds^2&=e^a e^a,\quad e^a=\sqrt{k(1-\lambda^4)}(D-\lambda^2)^{-1}_{ab}L^b.
\eea
Expressions for $D$ and $L$ are given in (\ref{ActDeform}).

\bigskip
\noindent
{\bf Dilaton for the $\la$--deformation}

Although extraction of the metric and the Kalb--Ramond field for the lambda deformation is rather straightforward, the procedure for calculating the dilaton is controversial. The original proposal of \cite{SftsGr} suggested the expression
\bea\label{BosDilatonGr}
e^{-2\Phi_B}=e^{-2\Phi_0} k^{\mbox{dim} G}\mbox{det} (\lambda^{-2}-D),
\eea
which can be written as 
\bea\label{BosDilaton1Gr}
e^{2\Phi_B}=\frac{1}{\mbox{det}[(Ad_f-\lambda^{-2})|_{\hat{f}}]},
\eea
where the determinant is taken in the algebra. In \cite{HMS} it was argued that for supergroups and supercosets an alternative expression is more appropriate:
\bea\label{SuDilatonGr}
e^{2\Phi}=\frac{1}{\mbox{sdet}[(Ad_f-\lambda^{-2})|_{\hat{f}}]},
\eea
Here the superdeterminant is computed in the full superalgebra $\hat{f}$. The difference between (\ref{BosDilaton1Gr}) and (\ref{SuDilatonGr}) originates from difference in the gauge fields which have been integrated out.

Recalling that an element of a superalgebra can be written as 
\bea\label{SuCosetDef} 
{\cal M}=\left[\begin{array}{c|c}
A&B\\
\hline
C&D
\end{array}
\right],
\eea
where $(A,D)$ are even and $(B,C)$ are odd blocks \cite{ArutFrolov,Beisert}, the expression (\ref{SuDilatonGr}) becomes
\bea\label{SuDilaton1}
e^{2\Phi}=\frac{\mbox{det}[(Ad_f-\lambda^{-2})|_{\hat{f}_1\oplus \hat{f}_3}]}{\mbox{det}[(Ad_f-\lambda^{-2})|_{\hat{f}_0\oplus \hat{f}_2}]}.
\eea
Here $\hat{f}_0$ and  $\hat{f}_2$ refer to the  even subspaces $A$ and $D$, while $\hat{f}_1$ and  $\hat{f}_3$ 
refer to the odd subspaces $B$ and $C$. In this article we will refer to (\ref{BosDilaton1Gr}) (which is equal to the denominator of (\ref{SuDilaton1})) as the {\it bosonic prescription}, and the numerator of (\ref{SuDilaton1}) would be called the {\it fermionic contribution} to the dilaton.

\bigskip
\noindent
{\bf $\la$--deformation for cosets}

The extension of the $\la$--deformation to cosets $G/H$ is presented in \cite{ST}. Separating the generators $T^A$ of $G$ into $T^a$ corresponding to $H \subset G$ and $T^\alpha$ corresponding to the coset $G/H$, one finds the metric
\bea\label{MainCoset}
ds^2&=&e^\alpha e^\alpha,\quad e^\alpha=-\sqrt{\frac{k(1-\lambda^4)}{2\lambda^4}}(M^{-1})^{\alpha B}L^B,\nn
M_{AB}&=&\begin{bmatrix} (D-1)_{ab} & D_{a \beta}\\ D_{\alpha b} & (D-\lambda^{-2}1)_{\alpha \beta} \end{bmatrix}, \quad D_{AB}=\mbox{Tr}(T_A g^{-1} T_B g),\\
L^A&=&-i\mbox{Tr}(g^{-1}dg T^A),\quad \mbox{Tr}(T_A T_B)=\delta_{AB}.\nonumber
\eea
The expression for the dilaton is given by the generalizations of (\ref{BosDilaton1Gr}) and (\ref{SuDilaton1}) \cite{ST,HMS,HT}:
\bea\label{BosDilaton1}
e^{2\Phi_B}&=&\frac{1}{\mbox{det}[(Ad_f-1-(\lambda^{-2}-1)P_\lambda)]},\\
\label{SuDilaton}
e^{2\Phi}&=&\frac{\mbox{det}[(Ad_f-1-(\lambda^{-2}-1)P_\lambda)|_{\hat{f}_1\oplus \hat{f}_3}]}{\mbox{det}[(Ad_f-1-(\lambda^{-2}-1)P_\lambda)|_{\hat{f}_0\oplus \hat{f}_2}]}.
\eea
Here $P_\la$ is a projector which separates the generators of $H$ and the coset $G/H$, and it has the 
form \cite{HMS}
\bea\label{PlaDef}
P_\la=P_2+\frac{\la}{\la+1}[P_1-\la P_3],\quad P_1+P_3=1.
\eea
Here $P_2$ is the projector in the bosonic sector, which can be written as
\bea
P_2=\begin{bmatrix} 0_{ab} & 0_{a \beta}\\ 0_{\alpha b} & 1_{\alpha \beta} \end{bmatrix}.
\eea
The action of fermionic projectors $P_1$ and $P_3$ is evaluated on a case--by--case basis, and we will address this question in the sections \ref{SecAdS3} and \ref{SectAdS5}. 
Notice that $P_2$ has already appeared in the matrix $M$ defined in (\ref{MainCoset}):
\bea
M_{AB}=D_{AB}-1-(\la^{-2}-1)P_2=Ad_f-1-(\lambda^{-2}-1)P_2.
\eea

We conclude this discussion with reviewing a very interesting observation made in \cite{SfetsosAdS5}: factorization of the $\la$--dependence in the determinant of $M_{AB}$. This technical simplification becomes especially useful in the AdS$_5\times$S$^5$ case, where one has to deal with large matrices. Following \cite{SfetsosAdS5}, we write $M_{AB}$ as a product of two block--triangular matrices: 
\bea\label{TriangM}
M=\left[\begin{array}{cc}
{\bf A}&0\\
{\bf C}&{\bf I}
\end{array}\right]\left[\begin{array}{cc}
{\bf I}&{\bf A}^{-1}{\bf B}\\
0&{\bf P}
\end{array}\right]\,.
\eea
As demonstrated in \cite{SfetsosAdS5}, matrix ${\bf P}$ has eigenvalues $\la^{-2}\pm 1$, so the coordinate dependence of the bosonic dilaton (\ref{BosDilaton1}) comes from $\mbox{det}\,{\bf A}$. We find that direct evaluation of the determinant of $M$ is easier than construction of ${\bf P}$, but our final results confirm that the coordinate dependence of $\mbox{det}\, M$ is inherited from $\mbox{det}\, {\bf A}$. 


\section{Deformation of AdS$_3\times$S$^3$}
\label{SecAdS3}

\renewcommand{\theequation}{3.\arabic{equation}}
\setcounter{equation}{0}

Let us apply the procedure reviewed in the last section to  AdS$_3\times$S$^3$. The bosonic part of the sigma model is described by a product of two cosets
\bea\label{BosCosetAdS3}
\frac{SU(2)\times SU(2)}{SU(2)_{diag}}\times \frac{SU(1,1)\times SU(1,1)}{SU(1,1)_{diag}},
\eea
and the full string theory is described by a super--coset \cite{AdS3PSU}\footnote{Various aspects of integrability of string on this background are further discussed in \cite{AdS3Int}.}
\bea\label{SuCosetAdS3}
\frac{PSU(1,1|2)^2}{SU(1,1)\times SU(2)}\,.
\eea
In section \ref{SecAdS3metr} we construct the metric and the bosonic contribution to the dilaton for the cosets (\ref{BosCosetAdS3}), (\ref{SuCosetAdS3}). While this will give the full answer for (\ref{BosCosetAdS3}), the dilaton for the supercoset (\ref{SuCosetAdS3}) also receives a fermionic contribution, which will be evaluated in section \ref{SecAdS3ferm}. In section \ref{SecAdS3RR} we construct the Ramond--Ramond fluxes supporting the $\la$--deformed supercoset (\ref{SuCosetAdS3}), and properties of the new geometries are discussed in sections \ref{SecAdS3SpecCases} and  \ref{SecAltParamAdS3}.

\subsection{Metric and the bosonic dilaton}
\label{SecAdS3metr}

The metric is constructed using the bosonic coset (\ref{BosCosetAdS3}), then $S^3$ and $AdS_3$ decouple, and they can be studied separately. We begin with analyzing the sphere, and deformation of $AdS_3$ can be found by performing an analytic continuation. 

\noindent
{\bf Deformation of the sphere.}

To describe the coset $\frac{SU(2)_l\times SU(2)_r}{SU(2)_{diag}}$, we use the algebraic parameterization introduced in \cite{ST}:
\bea\label{GandGprime}
g_l=\left[\begin{array}{cc}
\alpha_0+i\alpha_3&\alpha_2+i\alpha_1\\
-\alpha_2+i\alpha_1&\alpha_0-i\alpha_3
\end{array}\right],\qquad
g_r=\left[\begin{array}{cc}
\beta_0+i\beta_3&\beta_2+i\beta_1\\
-\beta_2+i\beta_1&\beta_0-i\beta_3
\end{array}\right]\,,
\eea
where variables $\alpha_k$, $\beta_k$ are subject to the determinant constraints
\bea\label{DetConstrt}
\sum (\alpha_k)^2=1,\qquad \sum (\beta_k)^2=1.
\eea
Gauging of the diagonal part of $SU(2)_l\times SU(2)_r$ makes the description (\ref{GandGprime}) redundant, and to remove the unphysical degrees of freedom we impose a convenient gauge, which was also used in \cite{ST}. Acting on $g_l$ as $g_l\rightarrow h^{-1}g_l h$, we can set $\alpha_2=\alpha_3=0$, then the remaining $U(1)$ transformations $h=\exp[ix \sigma_1]$ can be used to set $\beta_3=0$:
\bea\label{GaugeFixedS3}
\alpha_2=\alpha_3=\beta_3=0.
\eea
Following \cite{ST} we introduce a convenient coordinate $\gamma$ and solve the constraints (\ref{DetConstrt}) to express all remaining components of  $g_1$ and $g_2$  in terms of $(\alpha_0,\beta_0,\gamma)$:
\bea\label{OurVariablesAdS3}
 \beta_1\equiv \frac{\gamma}{\sqrt{1-\alpha_0^2}},\qquad\alpha_1=\sqrt{1-\alpha_0^2},\quad \beta_2=\sqrt{1-\beta_0^2-\frac{\gamma^2}{{1-\alpha_0^2}}}\,.
\eea
To simplify notation, we will drop the subscripts of ${\alpha_0}$ and $\beta_0$. 

The elements of $SU(2)_l\times SU(2)_r$ can be represented as block--diagonal $4\times 4$ matrices:
\bea\label{CosetElementS3}
g=\left(\begin{array}{cc}
g_l&0\\ 0&g_r
\end{array}\right),\qquad g^\dagger g=I,
\eea
then the generators corresponding to the subgroup $H$ and to the coset $G/H$ can be written in terms of the Pauli matrices\footnote{Recall that construction (\ref{MainCoset}) is based on normalized generators, and factor $1/2$ in \eqref{GeneratorsS3} ensures that $\mbox{Tr}(T_AT_B)=\delta_{AB}$.}.
\bea\label{GeneratorsS3}
H=SU(2)_{diag}:&&T_a=\frac{1}{2}\left[\begin{array}{cc}
\sigma_a &0\\ 0&\sigma_a
\end{array}\right],\quad a=1,2,3;\nn
G/H=\frac{SU(2)_l\times SU(2)_r}{SU(2)_{diag}}:&&T_\alpha=\frac{1}{2}\left[\begin{array}{cc}
\sigma_{\alpha-3} &0\\ 0&-\sigma_{\alpha-3}
\end{array}\right],\quad \alpha=4,5,6.
\eea
Substitution of \eqref{CosetElementS3}--\eqref{GeneratorsS3}, where $g_l,g_r$ are given by \eqref{GandGprime}, \eqref{GaugeFixedS3}, into the defining relations (\ref{MainCoset})\footnote{Recall  the ranges of indices in \eqref{MainCoset}: $a=\{1,2,3\}$, $\alpha=\{3,5,6\}$, $B=\{1,...,6\}$.} leads to the metric \cite{ST}
\bea\label{S3metr}
ds^2&=&\frac{k}{2(1-\lambda^4)\Lambda}\Delta_{\mu\nu} dx^\nu dx^\nu ,\qquad 
\Lambda=(1-\alpha^2)(1-\beta^2)-\gamma^2,\nn
\Delta_{\alpha\alpha}&=&4(1+\lambda^2)^2-\beta^2(3+\lambda^2)(1+3\lambda^2),\qquad
\Delta_{\alpha\gamma}=-\beta(1-\lambda^2)^2\\
\Delta_{\beta\beta}&=&4(1+\lambda^2)^2-\alpha^2(3+\lambda^2)(1+3\lambda^2),\qquad
\Delta_{\beta\gamma}=-\alpha(1-\lambda^2)^2
\nn
\Delta_{\gamma\gamma}&=&(1-\lambda^2)^2,\qquad
\Delta_{\alpha\beta}=\alpha\beta(1-\lambda^2)^2+4\gamma (1+\lambda^2)^2.\nonumber
\eea

\noindent
{\bf Deformation of AdS$_3$.}

The deformation of the AdS$_3$ is constructed by performing an analytic continuation of \eqref{S3metr}. 
The defining relation for $g\in SU(1,1)_l\times SU(1,1)_r$ is 
\bea\label{gAdSFull}
g=\left[\begin{array}{cc}
g_l&0\\ 0&g_r
\end{array}\right],\quad
g^{\dagger}\Sigma_4g=\Sigma_4,\quad 
\Sigma_4=\begin{bmatrix}
1&0&0&0\\
0&-1&0&0\\
0&0&1&0\\
0&0&0&-1
 \end{bmatrix},
\eea
and it can be enforced by starting with an element of $SU(2)_l\times SU(2)_r$, renaming the coordinates as 
\bea\label{AnalContAdS3}
\alpha\to \tilde{\alpha},\quad \beta\to \tilde{\beta},\quad \gamma\to \tilde{\gamma},\quad k\rightarrow -k,
\eea
and changing their range from 
\bea\label{Range1}
0<\alpha^2<1,\quad 0<\beta^2<1,\quad \gamma^2<(1-\alpha^2)(1-\beta^2)
\eea
to
\bea\label{Range2}
1<\tilde{\alpha}^2,\quad 1<\tilde{\beta}^2,\quad \tilde{\gamma}^2<(\tilde{\alpha}^2-1)(\tilde{\beta}^2-1).
\eea
To view this transition as a proper analytic continuation, one can introduce alternative coordinates $(a,b,\gamma)$ as
\bea
a^2=1-\alpha^2,\quad b^2=1-\beta^2.
\eea
Then transition from (\ref{Range1}) to (\ref{Range2}) amounts to a continuation from real to imaginary $(a,b)$. This changes the signature from $(+++)$ to $(--+)$, and by changing the sign of $k$ we recover $(++-)$. 

Analytic continuation (\ref{AnalContAdS3}) along with the replacement $k\to -k$ gives the metric for the $\lambda$--deformed $AdS_3$
\bea\label{AdS3metr}
d\tilde{s}^2&=&\frac{k}{2(1-\lambda^4)\tilde{\Lambda}}\tilde{\Delta}_{\mu\nu} dx^\nu dx^\nu ,\qquad \tilde{\Lambda}=(\tilde{\alpha}^2-1)(\tilde{\beta}^2-1)-\tilde{\gamma}^2,\nn
\tilde{\Delta}_{\tilde{\alpha}\tilde{\alpha}}&=&-4(1+\lambda^2)^2+\tilde{\beta}^2(3+\lambda^2)(1+3\lambda^2),\qquad
\tilde{\Delta}_{\tilde{\alpha}\tilde{\gamma}}=\tilde{\beta}(1-\lambda^2)^2\\
\tilde{\Delta}_{\tilde{\beta}\tilde{\beta}}&=&-4(1+\lambda^2)^2+\tilde{\alpha}^2(3+\lambda^2)(1+3\lambda^2),\qquad
\tilde{\Delta}_{\tilde{\beta}\tilde{\gamma}}=\tilde{\alpha}(1-\lambda^2)^2
\nn
\tilde{\Delta}_{\tilde{\gamma}\tilde{\gamma}}&=&-(1-\lambda^2)^2,\qquad
\tilde{\Delta}_{\tilde{\alpha}\tilde{\beta}}=-\tilde{\alpha}\tilde{\beta}(1-\lambda^2)^2-4\tilde{\gamma} (1+\lambda^2)^2.\nonumber
\eea

\noindent
{\bf Dilaton and RR fields for the bosonic coset.}

The deformation of AdS$_3\times$S$^3$ constructed in \cite{ST} is described by the metric $\{$(\ref{S3metr}), (\ref{AdS3metr})$\}$ and the dilaton corresponding to the {\it bosonic} prescription (\ref{BosDilaton1}):
\bea
e^{-2\Phi_B}=\frac{2\Lambda (1-\lambda^2)^2(1+\lambda^2)}{\lambda^6}\,
\frac{2\tilde{\Lambda}(1-\lambda^2)^2(1+\lambda^2)}{\lambda^6}=e^{-2\Phi_0}\Lambda \tilde{\Lambda},
\eea
This article also listed the corresponding Ramond--Ramond fields:
\bea\label{C2Bos}
C_2=
\frac{4k\lambda}{1-\lambda^2}
\left[\tilde{\beta} \beta d\tilde{\alpha}\wedge d\alpha+2\tilde{\beta} \alpha d\tilde{\alpha}\wedge d\beta-\tilde{\beta}d\tilde{\alpha}\wedge d\gamma+\tilde{\alpha}\alpha d\tilde{\beta}\wedge d\beta-\alpha d\tilde{\gamma}\wedge d\beta\right].
\eea
However, as argued in \cite{HMS,HT,BTW}, the dilaton (\ref{SuDilaton}) for the supercoset is more natural for describing superstrings, and in the next subsection we will find the appropriate expression and construct the corresponding Ramond--Ramond fluxes.

\subsection{Fermionic contribution to the dilaton}
\label{SecAdS3ferm}

In this subsection we will construct the dilaton for the supercoset (\ref{SuCosetAdS3}) using the prescription (\ref{SuDilaton}). Before focusing on (\ref{SuCosetAdS3}), we will outline the procedure for applying (\ref{SuDilaton}) to a supermatrix (\ref{SuCosetDef}) constructed from extending an algebra of the bosonic coset $(G_1/H_1)\times (G_2/H_2)$. 

A supersymmetric extension of av algebra ${\mathfrak g}_1\times {\mathfrak g}_2$ has the form
\bea\label{SuGenerator}
{\cal M}=\left[\begin{array}{cc}
{\tgt{g}}_{1}&{\tgt{f}}_{12}\\ {\tgt{f}}_{21}&{\tgt{g}}_{2}
\end{array}\right],\qquad {\tgt{g}}_{1}\in {\mathfrak g}_1,\quad {\tgt{g}}_{2}\in {\mathfrak g}_2,
\eea
and to find the supercoset, we should fix the gauge corresponding to subalgebras ${\mathfrak h}_1$, 
${\mathfrak h}_2$ and evaluate the relevant projectors  $P_\lambda$. This can be done in five steps:
\begin{enumerate}
\item Find an automorphism $J_1$ of algebra ${\mathfrak g}_1$ which leaves invariant only the elements of 
${\mathfrak h}_1$. In other words, $g\in {\mathfrak g}_1$ satisfies the condition
\bea\label{JprojBos}
J_1^{-1} {g}J_1={g}
\eea
if and only if ${g}\in {\mathfrak h}_1$. Automorphism $J_2$ in ${\mathfrak g}_2$ is defined in a similar way. 
\item
Construct an automorphism of the super-algebra as
\bea
{\cal P}=\left[\begin{array}{cc}
J_1&0\\ 0&J_2
\end{array}\right],
\eea
and project out the elements ${\cal M}$ which are left invariant under such automorphism\footnote{Sometimes this condition requires a modification: as we will see in section \ref{SectAdS5}, in the case of AdS$_5\times$S$^5$ it must be replaced by ${\cal P}^{-1}{\cal M}{\cal P}={\cal M}^T$.}: 
\bea\label{SuProj}
{\cal P}^{-1}{\cal M}{\cal P}={\cal M}\,.
\eea
For bosonic generators this reduces to (\ref{JprojBos}) and its counterpart for ${\mathfrak g}_2$, while the projections for the  fermionic matrices are
\bea\label{FermProj}
J_1^{-1}{\tgt{f}}_{12}J_2={\tgt{f}}_{12},\qquad J_2^{-1}{\tgt{f}}_{21}J_1={\tgt{f}}_{21}.
\eea
\item Construct the projector $P_2$ acting on bosonic generators by requiring that $[1-P_2]$ kills the same elements as (\ref{JprojBos}) and its counterpart with $J_2$. Such $P_2$ projects on the bosonic part of the supercoset.
\item Construct projector $P_3$ acting on fermionic generators by requiring that $P_3$ keeps the same elements as (\ref{FermProj}). The fermionic projector complementary to $P_3$ is $P_1=1-P_3$.
\item Construct the projector $P_\la$ using the definition (\ref{PlaDef}). Substitution of this expression into (\ref{SuDilaton}) or (\ref{BosDilaton1}) and evaluation  of the resulting determinant gives the dilaton for the (super)coset. 
\end{enumerate}
To apply this procedure to the $AdS_3 \times S^3$ coset (\ref{SuCosetAdS3}), we observe that 
${\mathfrak g}_1$ represents the algebra of (\ref{CosetElementS3}),
\bea
g\in {\mathfrak g}_1:\quad g=\left(\begin{array}{cc}
g_l&0\\ 0&g_r
\end{array}\right),\quad g_l\in {\mathfrak{su}}(2),\ g_r\in {\mathfrak{su}}(2),
\eea
while the elements of 
${\mathfrak h}_1={\mathfrak{su}}(2)_{diag}$ have the form
\bea
\left[\begin{array}{cc}
g&0\\ 0&g
\end{array}\right],\qquad g\in {\mathfrak{su}}(2).
\eea
This leads to two options for the automorphism $J_1$:
\bea
J_1=\pm\left[\begin{array}{cc}
0&1_{2\times 2}\\ 1_{2\times 2}&0
\end{array}\right]\,.
\eea
Expression for $J_2$ is constructed in a similar way, and putting these results together, we find two options for the automorphism ${\cal P}$:
\bea\label{CalPforAdS3}
{\cal P}=\left[\begin{array}{cccc}
0&1_{2\times 2}&0&0\\ 
1_{2\times 2}&0&0&0\\
0&0&0&1_{2\times 2}\\ 
0&0&1_{2\times 2}&0
\end{array}\right]\quad\mbox{or}\quad
{\cal P}=\left[\begin{array}{cccc}
0&1_{2\times 2}&0&0\\ 
1_{2\times 2}&0&0&0\\
0&0&0&-1_{2\times 2}\\ 
0&0&-1_{2\times 2}&0
\end{array}\right]\,.
\eea
The fermionic generators of $PSU(1,1|2)\times PSU(1,1|2)$ appearing in (\ref{SuGenerator}) obey the relation 
\bea
f_{12}=-i\Sigma_4 (f_{21})^\dagger
\eea
with $\Sigma_4$ given in (\ref{gAdSFull}), and projection (\ref{SuProj}) leads to further constraints. It is convenient to decouple $f_{12}$ and $f_{21}$ by working with holomorphic and anti--holomorphic coordinates. Relations (\ref{FermProj}) isolate $4+4$ components of $f_{12}$ and $f_{21}$ killed by $P_1$, while $P_3$ kills the complementary $4+4$ components\footnote{Recall that even though $f_{12}$ and $f_{21}$ are represented by $4\times 4$ matrices, each of these objects has only $8$ nonzero components. The details are discussed in the Appendix \ref{AppPSU}, here we just refer to the explicit form of the $\mathfrak{psu}(1,1|2)\times\mathfrak{psu}(1,1|2)$ matrix (\ref{MAdS3Split}), which clearly exhibits the non--vanishing elements.}. Extraction of $(P_1,P_2,P_3)$, construction of $P_\la$ via (\ref{PlaDef}), and evaluation of superdeterminant  (\ref{SuDilaton}) gives the same dilaton for both choices (\ref{CalPforAdS3}):
\bea\label{SuDilatonAdS3}
e^{\Phi}&=&Qe^{\Phi_B},\quad e^{\Phi_B}=\frac{1}{\sqrt{[(1-\alpha^2)(1-\beta^2)-\gamma^2][(\tilde{\alpha}^2-1)(\tilde{\beta}^2-1)-\tilde{\gamma}^2]}},\nn
Q
&=&(1-\lambda^2)^4\Big[\gamma+\tilde{\gamma}-\frac{4\la(1+\la^2)}{(1-\la^2)^2}(\alpha{\tilde\beta}+\tilde{\alpha}\beta) +
\frac{\la^4+6\la^2+1}{(\la^2-1)^2}(\alpha\beta+\tilde{\alpha}\tilde{\beta} )\Big]^2\,.
\eea

We conclude this section by analyzing the symmetries of the metric $\{$(\ref{S3metr}), (\ref{AdS3metr})$\}$ and the dilaton (\ref{SuDilatonAdS3}), which will be used for constructing the Ramond--Ramond fluxes. First, it is clear that neither the metric nor the dilaton has continuous symmetries, but all NS--NS fluxes are invariant under several discrete transformations:
\bea\label{DicreteSymm}
S_1:&&\alpha\leftrightarrow \beta,\qquad {\tilde\alpha}\leftrightarrow{\tilde\beta}\,;\nn
S_2:&&\alpha\leftrightarrow {\tilde\alpha},\qquad {\beta}\leftrightarrow{\tilde\beta},\quad\gamma\leftrightarrow{\tilde\beta},\quad
k\leftrightarrow(-k)\,.
\eea
These symmetries will be used in the next section to select a natural solution for the RR field $C_2$.

\subsection{Ramond--Ramond fluxes}
\label{SecAdS3RR}

Although the Ramond--Ramond fluxes for the lambda--deformed backgrounds can be extracted from the fermionic part of the sigma model, such problem is notoriously complicated \cite{BTW}. When similar deformation were analyzed in the past, the RR fluxes were obtained by solving supergravity equations \cite{OtherEta, LRT,BTW}, and in this section we will follow the same route. We will demonstrate that under very weak assumptions, supergravity gives the unique expression for all fluxes. 

Since the undeformed AdS$_3\times$S$^3$ geometry is supported by the Ramond--Ramond three-form, we assume that the situation will remain the same after the deformation, so the relevant part of action for the type IIB supergravity reads
\bea
S=\int d^6 x\sqrt{-g}\left[ e^{2\Phi}(R+4(\d \Phi)^2) - \frac{1}{12}F_{mnp}F^{mnp} \right].
\eea
This leads to the equations of motion
\bea\label{EOM1}
&&\nabla^2 e^{-2\Phi}=0,\\
\label{EOM2}
&&\nabla_m F^{mnk}=0, \\
\label{EOM3}
&&e^{-2\Phi}(R_{mn}+2\nabla_m\nabla_n \Phi)=\frac{1}{4}\left( F_{mpq}F_n{}^{pq}-\frac{1}{6}g_{mn}F_{spq}F^{spq} \right)
\eea
and the first one is solved by metric (\ref{S3metr}), (\ref{AdS3metr}) and the dilaton (\ref{SuDilatonAdS3}).

To construct an expression for $C_2$, we observe that the left--hand side of the Einstein's equation (\ref{EOM3}) has the structure
\bea
\frac{P}{Q^2},
\eea
where  $Q$ is given by (\ref{SuDilatonAdS3}), and $P$ is a polynomial in $(\alpha,\beta,\gamma,{\tilde\alpha},{\tilde\beta},{\tilde\gamma})$. This suggests a natural ansatz for $C_2$:
\bea
C_2=\frac{1}{Q}{\tilde C}_{\mu\nu}dx^\mu\wedge dx^\nu,
\eea
where all ${\tilde C}_{\mu\nu}$ are polynomials of degree two\footnote{The degree comes from counting powers in the left--hand side of the Einstein's equations.} in 
$(\alpha,\beta,\gamma,{\tilde\alpha},{\tilde\beta},{\tilde\gamma})$. This ansatz leaves 
\bea
\frac{6\times 5}{2}\times \left[1+6+6+\frac{6\times 5}{2}\right]=420
\eea
undetermined coefficients. We then found the most general solution for ${\tilde C}_{\mu\nu}$ following these steps:
\begin{enumerate}
\item Solving equations (\ref{EOM2})--(\ref{EOM3}) for $\la=0$, when the metric and the dilaton are relatively simple, we reduced the number of undetermined coefficients to 43.
\item Solving equations (\ref{EOM2})--(\ref{EOM3})  in the first order in $\la$, we reduced the number of undetermined coefficients in the {\it zeroth} order to 42.
\item Eliminating the gauge freedom, we demonstrated that the solution at the zeroth order in $\la$ is {\it unique} up to a gauge transformation.
\end{enumerate}
Once uniqueness of the solution for $\la=0$ is demonstrated, we can choose a convenient gauge which respects the discrete symmetries (\ref{DicreteSymm}):
\bea
&&C_{\alpha{\tilde\alpha}}=\frac{k}{Q}\left[2-(\beta^2+{\tilde\beta}^2)\right],\quad
C_{\beta{\tilde\beta}}=-\frac{k}{Q}\left[2-(\alpha^2+{\tilde\alpha}^2)\right],\quad\\
\quad
&&C_{\alpha{\tilde\beta}}=-C_{\beta{\tilde\alpha}}=k\frac{{\tilde\gamma}-\gamma}{Q},\quad
C_{\alpha{\tilde\gamma}}=-\frac{k{\tilde\beta}}{Q},\quad
C_{\beta{\tilde\gamma}}=\frac{k{\tilde\alpha}}{Q},\quad
C_{\gamma{\tilde\alpha}}=-\frac{k{\beta}}{Q},\quad C_{\gamma{\tilde\beta}}=\frac{k{\alpha}}{Q}\,.
\nonumber
\eea
This solution is odd under $S_1$ and $S_2$. The uniqueness of the solution in the zeroth order in $\la$ guarantees 
that, up to a gauge transformation, there is a unique gauge potential $C_2$, at least in the perturbative expansion in powers of $\la$. Making a guess consistent with symmetries (\ref{DicreteSymm}), we arrive at the final solution 
\bea\label{C2forAdS3}
&&C_{\alpha{\tilde\alpha}}=\frac{\hat k}{Q}\left[2+c_1\beta{\tilde\beta}-c_3(\beta^2+{\tilde\beta}^2)\right],\quad
\quad
C_{\alpha{\tilde\beta}}=-C_{\beta{\tilde\alpha}}=\frac{\hat k}{Q}({\tilde\gamma}-\gamma),\nn
&&C_{\alpha{\tilde\gamma}}=\frac{\hat k}{Q}[c_2\beta-{\tilde\beta}]\quad
C_{\beta{\tilde\beta}}=-\frac{\hat k}{Q}\left[2+c_1\alpha{\tilde\alpha}-c_3(\alpha^2+{\tilde\alpha}^2)\right],\quad
C_{\beta{\tilde\gamma}}=-\frac{\hat k}{Q}[c_2\alpha-{\tilde\alpha}],\nn
&&C_{\gamma{\tilde\alpha}}=-\frac{\hat k}{Q}[{\beta}-c_2{\tilde\beta}],\qquad 
C_{\gamma{\tilde\beta}}=\frac{\hat k}{Q}[{\alpha}-c_2{\tilde\alpha}]
\\
&&c_1=2c_2c_3,\qquad c_2=\frac{2\la}{1+\la^2},\qquad c_3=\frac{\la^4+6\la^2+1}{(\la^2-1)^2},\quad {\hat k}=\frac{k(1+\lambda^2)}{1-\lambda^2}\,.\nonumber
\eea
Notice that, unlike the solution (\ref{C2Bos}) with the ``bosonic dilaton'', the field (\ref{C2forAdS3}) has a  complicated lambda dependence, and the situation is similar in the AdS$_2\times$S$^2$ case, which is reviewed in the Appendix \ref{AppAdS2}. In particular, while the field  (\ref{C2Bos}) vanishes at the WZW point ($\la=0$), our solution for the supercoset (\ref{C2forAdS3}) goes to a nontrivial limit, and, as we will see in section \ref{SectAdS5} and in the Appendix \ref{AppAdS2}, the same phenomenon persists for AdS$_2\times$S$^2$ and AdS$_5\times$S$^5$.

\bigskip

To summarize, the $\la$--deformed version of AdS$_3\times$S$^3$ is described by the metric (\ref{S3metr}), (\ref{AdS3metr}), the dilaton (\ref{SuDilatonAdS3}), and the Ramond--Ramond two--form (\ref{C2forAdS3}). In the next subsection we will analyze some special cases of this geometry.

\subsection{Special cases}
\label{SecAdS3SpecCases}

The solution (\ref{S3metr}), (\ref{AdS3metr}), (\ref{SuDilatonAdS3}), (\ref{C2forAdS3}) simplifies in several special cases, and we will briefly discuss these interesting limits.

The gauged WZW model is obtained by setting $\la=0$: 
\bea\label{WZWsss}
ds^2&=&\frac{k}{2\Lambda}\left[4(1-\beta^2)d\alpha^2+4(1-\alpha^2)d\beta^2+8\gamma d\alpha d\beta+(d\gamma-\beta d\alpha-\alpha d\beta)^2\right]+\nn
&&+\frac{k}{2\Lambda}\left[4({\tilde\beta}^2-1)d{\tilde\alpha}^2+4({\tilde\alpha}^2-1)d{\tilde\beta}^2-
8{\tilde\gamma} d{\tilde\alpha} d{\tilde\beta}+(d{\tilde\gamma}-{\tilde\beta} d{\tilde\alpha}-
{\tilde\alpha} d{\tilde\beta})^2\right]\,,\nn
\qquad 
\Lambda&=&(1-\alpha^2)(1-\beta^2)-\gamma^2,\qquad 
\tilde{\Lambda}=(\tilde{\alpha}^2-1)(\tilde{\beta}^2-1)-\tilde{\gamma}^2\,,\\
e^{\Phi}&=&\frac{Q}{\sqrt{\Lambda{\tilde\Lambda}}}e^{\Phi_B},\quad
Q
=\Big[\gamma+\tilde{\gamma}+\alpha\beta+\tilde{\alpha}\tilde{\beta} \Big]^2\,,\nn
C_2&=&\frac{k}{Q}\left[({\tilde\alpha}d\beta-{\tilde\beta}d\alpha)\wedge (d{\tilde\gamma}+{\tilde\alpha}d{\tilde\beta}+{\tilde\beta}d{\tilde\alpha})-
({\alpha}d{\tilde\beta}-{\beta}d{\tilde\alpha})\wedge (d{\gamma}+{\alpha}d{\beta}+{\beta}d{\alpha})
\right.\nn
&&\left.+({\tilde\gamma}-{\gamma}+{\tilde\alpha}\tilde\beta-\alpha{\beta})(d\alpha\wedge d{\tilde\beta}-d\beta\wedge d{\tilde\alpha})+2(d\alpha\wedge d{\tilde\alpha}-d\beta\wedge d{\tilde\beta})
\right]\,.\nonumber
\eea
This should be contrasted with bosonic gWZW, which has the dilaton
\bea
e^{\Phi}&=&\frac{1}{\sqrt{\Lambda{\tilde\Lambda}}}
\eea
and vanishing $C_2$ (see (\ref{C2Bos})). A similar contrast is encountered in the AdS$_2\times$S$^2$ and AdS$_5\times$S$^5$ cases, which discussed in section \ref{SectAdS5} and in the Appendix \ref{AppAdS2}.

Note that the metric (\ref{WZWsss}) for the $SO(4)/SO(3)$ gWZW model has been discussed in 
\cite{Fradkin,MoreEta}, where the element of the coset was defined as
\bea\label{Fradkin}
g&=&g_1(\varphi)g_2(\theta)g_3(2t)g_2(\theta)g_1(\varphi),\\
g_k(\alpha)&=&\exp(\alpha T_{k,k+1}),\quad (T_{k,k+1})_i^j=\delta_{k,i}\delta_{k+1}^j-\delta_{k+1,i}\delta_k^j,\quad k=1,2,3.\nonumber
\eea
The coordinates used in (\ref{WZWsss}) are related to the Euler angles (\ref{Fradkin}) as
\bea
\alpha&=&\cos\varphi\cos t\cos\theta+\sin\varphi\sin t,\nn
\beta&=&\cos\varphi\cos t\cos\theta-\sin\varphi\sin t,\\
\gamma&=&-\cos^2\varphi\sin^2 t+\cos^2 t(\cos^2\theta\sin^2\varphi+\sin^2\theta).\nonumber
\eea

\bigskip

Another interesting limit is obtained by setting $\la=1$. However, this limit should be approached with a great care since denominators contain $(\la^2-1)$. We will follow the procedure discussed in \cite{HT} adopting it to our coordinates. To arrive at a sensible limit, we rescale the coordinates on the sphere as 
\bea
\alpha\propto\frac{1}{\eps},\quad \beta\propto \frac{1}{\eps},\quad \gamma\propto\frac{1}{\eps^2}
\eea
and send $\eps$ to zero. This gives the metric of the $\eta$--deformed $S^3$ \cite{OtherEta}, and to see this, we introduce the standard coordinates 
$(r,\phi,\varphi)$ by 
\bea\label{KappaLimit}
\alpha&=&\frac{1}{\eps}\, \frac{e^{i(\varphi+\phi)}r}{(1+\lambda^2)\sqrt{2(1-r^2)}},\quad 
\beta=\frac{1}{\eps}\, \frac{e^{i(\varphi-\phi)}r}{(1+\lambda^2)\sqrt{2(1-r^2)}},\nn 
\gamma&=&\frac{1}{\eps^2}\, \frac{e^{2i\varphi}(2(1+\lambda^2)^2-(1-\lambda^2)^2r^2)}{2(1-\lambda^4)^2(1-r^2)},\quad \eps\to 0.
\eea
Performing a similar change of variables on AdS$_3$ along with an analytic continuation 
\bea
\phi\to\psi,\quad \varphi\to t,\quad r\to i\rho,\quad k\to-k,
\eea
and sending $\eps$ to zero, we arrive at the metric and the dilaton
\bea
ds^2&=&\frac{h}{2}\Big(\frac{1}{1-\varkappa^2 r^2}\left[(1-r^2)d\varphi^2+\frac{dr^2}{1-r^2} \right]+r^2d\phi^2\nn
&&+\frac{1}{1+\varkappa^2 \rho^2}\left[-(1+\rho^2)dt^2 + \frac{d\rho^2}{1+\rho^2}\right]+\rho^2d\psi^2\Big),\\
e^{\Phi}&=&(1+{\tilde\la}^2)^4\frac{\left[2(1-{\tilde\la}^2)S^2\cos(\varphi-t)-4{\tilde\la}\rho rS\cos(\phi-\psi)\right]^2}{
S^2\sqrt{(1-{\tilde\la}^2)^2+(1+{\tilde\la}^2)^2\rho^2}\sqrt{(1-{\tilde\la}^2)^2-(1+{\tilde\la}^2)^2r^2}},\nn
S&\equiv& \sqrt{(1+\rho^2)(1-r^2)},\quad \varkappa=\frac{1+{\tilde\la}^2}{1-{\tilde\la}^2},\quad 
h=\frac{(1-{\tilde\la}^2)^2}{k(1+{\tilde\la}^2)},\quad {\tilde\la}=i\la\,.
\nonumber
\eea 
This geometry describes the $\eta$--deformed AdS$_3\times$S$^3$ \cite{OtherEta}, and similar relations between $\la$-- and $\eta$--deformations have been explored in \cite{HT}.


\subsection{Alternative parameterizations}
\label{SecAltParamAdS3}

In subsections \ref{SecAdS3metr}--\ref{SecAdS3RR} we derived the full supergravity solutions corresponding to the $\la$--deformed supercoset, but the metric for this geometry has already appeared in the literature \cite{ST,HT}. We used the parameterization of \cite{ST}, and in this subsection we will discuss the relation with the coordinates used in \cite{HT} and discuss one more parameterization which becomes useful for comparing AdS$_3\times$S$^3$ and AdS$_5\times$S$^5$ solutions.

To find the relation between our parameterization and the coordinates used in \cite{HT}, we observe that the action by $H=SU(2)_{diag}$ changes components of $g_l$ and $g_r$ in (\ref{GandGprime}), but three expressions remain invariant:
\bea\label{VecInvar}
\vec{\alpha}^2\equiv \sum_{i=1}^3\alpha_i\alpha_i,\quad 
\vec{\beta}^2\equiv\sum_{i=1}^3\beta_i\beta_i,\quad 
\vec{\alpha}\cdot\vec{\beta}\equiv\sum_{i=1}^3\alpha_i\beta_i\,.
\eea
Although the gauge used in \cite{HT} was different from ours, we can find the map between two sets of coordinates by matching the expressions (\ref{VecInvar}) in two descriptions. The authors of \cite{HT} used parameterization in terms of the Euler's angles:
\bea\label{TrigParam}
g^{trig}=\exp[i\varphi \sigma_3\oplus (-\sigma_3)]\exp[i\zeta \sigma_1\oplus\sigma_1]\exp[i \phi \sigma_3\oplus\sigma_3].
\eea
Evaluating the invariants (\ref{VecInvar}) for parameterizations (\ref{GaugeFixedS3})--(\ref{OurVariablesAdS3}) and (\ref{TrigParam}), and comparing the results, we arrive at the map\footnote{Recall that to simplify notation we introduced $\alpha=\alpha_0$ and $\beta=\beta_0$, and all our results were written in these variables.}
\bea
\alpha=\cos(\varphi+\phi)\cos\zeta,\ \beta=\cos(\varphi-\phi)\cos\zeta,\
\gamma=\cos2\varphi-\frac{\cos2\varphi+\cos2\phi}{2}\cos^2\zeta.
\eea

Another interesting coordinate system comes from parameterizing the coset $SO(4)/SO(3)$ in terms of a three--dimensional vector $X$ and an anti--symmetric $3\times 3$ matrix $A$ \cite{Bars,SfetsosAdS5}. Such parameterization of $SO(n+1)/SO(n)$ will be used in the next section for studying the deformed AdS$_5\times$S$^5$, so it is important to introduce similar coordinates in the present case to make comparisons. The detailed discussion of parameterization and the gauge fixing is presented in section \ref{SectAdS5metr}, here we just write the result\footnote{We use variables $Y_1$ and $Y_2$ in (\ref{SO4gauge}) to make comparison with AdS$_5\times$S$^5$ case easier: the variable $Y_1$ is a counterpart of $X_1$, and $Y_2$ is a counterpart of $X_5$ in (\ref{SO6gauge}).}:
\bea\label{SO4gauge}
g&=&
\left[\begin{array}{cc}
1&0\\
0&(1+A)(1-A)^{-1}
\end{array}\right]
\left[\begin{array}{cccc}
b-1&bX_i\\
-bX_i&\delta_i^j-bX_iX^j
\end{array}\right]\,,\\
A&=&\left[\begin{array}{ccc}
0&a&0\\
-a&0&0\\
0&0&0
\end{array}\right],\quad 
b=\frac{2}{1+(Y_1)^2+(Y_2)^2},\quad {\vec X}=\{Y_1,0,Y_2\}\,.\nonumber
\eea 
The parameterizations (\ref{SO4gauge}) and (\ref{GandGprime}), (\ref{CosetElementS3}) correspond to different representations of $SO(4)$, so to relate them we should compare quantities which don't depend on the representation. We have already encountered such an object before:
\bea\label{DabMap}
D_{AB}=\mbox{Tr}(T_A g^{-1} T_B g).
\eea
To establish the map between generators, we recall that the subgroup $H=SU(2)_{diag}$ corresponds to
\bea
T_{H}^{SU(2)\times SU(2)}=
\frac{1}{2}\left[\begin{array}{cc}
x_a\sigma_a&0\\
0&x_a\sigma_a
\end{array}\right],\quad
T_{H}^{SO(4)}=\frac{i}{\sqrt{2}}\left[\begin{array}{cccc}
0&0&0&0\\
0&0&x_3&-x_2\\
0&-x_3&0&x_1\\
0&x_2&-x_1&0
\end{array}\right]
\eea
and the coset generators correspond to
\bea
T_{coset}^{SU(2)\times SU(2)}=
\frac{1}{2}\left[\begin{array}{cc}
y_a\sigma_a&0\\
0&-y_a\sigma_a
\end{array}\right],\quad
T_{coset}^{SO(4)}=\frac{i}{\sqrt{2}}\left[\begin{array}{cccc}
0&y_1&y_2&y_3\\
-y_1&0&0&0\\
-y_2&0&0&0\\
-y_3&0&0&0
\end{array}\right]
\eea
Evaluating (\ref{DabMap}) for (\ref{SO4gauge}) and $\{$(\ref{GandGprime}), (\ref{CosetElementS3})$\}$, using appropriate generators, and matching the results, we arrive at the map
\bea
&&\alpha=\frac{1-a Y_2}{\sqrt{1+a^2}Y},\quad \beta=\frac{1+a Y_2}{\sqrt{1+a^2}\,Y},\quad \gamma=-\frac{Y_1^2+a^2(Y_1^2-1)+Y_2^2}{(1+a^2)\,Y^2},\\
&&Y^2=1+(Y_1)^2+(Y_2)^2.\nonumber
\eea
and its inverse
\bea
&&a=-\frac{\sqrt{2(1+\gamma^2)-\alpha^2-\beta^2}}{\alpha+\beta},\quad Y_2=\frac{\alpha-\beta}{\sqrt{2(1+\gamma)-\alpha^2-\beta^2}},\nn
&&Y_1=-\sqrt{\frac{2[(1-\alpha^2)(1-\beta^2)-\gamma^2]}{[1+\alpha\beta+\gamma][2(1+\gamma)-(\alpha^2+\beta^2)]}}.
\eea
The AdS coordinates are obtained by the replacement
\bea
Y_1\to i {\tilde Y}_1,\quad Y_2\rightarrow Y_2,\quad a\rightarrow{\tilde a}.
\eea
In coordinates $(Y_1,Y_2,a,{\tilde Y}_1,{\tilde Y}_2,{\tilde a})$ the dilaton becomes
\bea
e^{\Phi}&=&Q e^{\Phi_B},\quad e^{\Phi_B}=\frac{\sqrt{1+a^2}Y \sqrt{1+\tilde{a}^2}\tilde{Y}}{16 a\tilde{a}Y_1\tilde{Y}_1},\quad Y^2=1+Y_1^2+Y_2^2,\quad \tilde{Y}^2=1-\tilde{Y}_1^2+\tilde{Y}_2^2,\nn
Q
&=&(1-\lambda^2)^4\Big[-\frac{Y_1^2+a^2(Y_1^2-1)+Y_2^2}{(1+a^2)Y^2}+\frac{\tilde{Y}_1^2+\tilde{a}^2(\tilde{Y}_1^2+1)+\tilde{Y}_2^2}{(1+\tilde{a}^2)\tilde{Y}^2}\\
&&-\frac{8\la(1+\la^2)}{(1-\la^2)^2}\frac{1-a\tilde{a}Y_2\tilde{Y}_2}{\sqrt{1+a^2}\sqrt{1+\tilde{a}^2}Y\tilde{Y}} 
+\frac{\la^4+6\la^2+1}{(\la^2-1)^2}\left(\frac{1-a^2 Y_2^2}{(1+a^2)Y^2}+\frac{1-\tilde{a}^2\tilde{Y}^2_2}{(1+\tilde{a}^2)\tilde{Y}^2} \right)\Big]^2.\nonumber
\eea
In particular, for the gauged WZW model ($\la=0$) we find
\bea\label{QforWZW3}
Q=4\left[\frac{X^2+{\tilde X}^2-X^2{\tilde X}^2}{X^2{\tilde X}^2}\right]^2\,.
\eea
Notice that this expression does not depend on coordinates $a$ and ${\tilde a}$, and the same phenomenon is encountered in the AdS$_5\times$S$^5$ case, see the last factor in (\ref{WZWforAdS5}).


\section{Towards the deformation of AdS$_5\times $S$^5$}

\label{SectAdS5}
\renewcommand{\theequation}{4.\arabic{equation}}
\setcounter{equation}{0}

In this section we apply the procedure described in section \ref{SecReview} to construct the $\lambda$--deformed AdS$_5\times $S$^5$ supercoset. Our final result includes the metric and the dilaton, but since the latter looks rather complicated, we were not able to solve the equations for the Ramond--Ramond fluxes. 

Superstrings on AdS$_5\times $S$^5$ are described by a  sigma model on the supercoset \cite{AdS5PSU}
\bea\label{AdS5suCoset}
\frac{PSU(2,2|4)}{SO(4,1)\times SO(5)}.
\eea
The corresponding superalgebra is represented by $4\times 4$ matrices, and an explicit parameterization is presented in the appendix \ref{AppPSU}. The bosonic part of the supercoset (\ref{AdS5suCoset}) is given by
\bea\label{AdS5BosCoset}
 \frac{SU(2,2)}{SO(4,1)}\times \frac{SU(4)}{SO(5)}= 
 \frac{SO(4,2)}{SO(4,1)}\times \frac{SO(6)}{SO(5)},
\eea 
and, as in the AdS$_3\times$S$^3$ case, the two subgroups decouple in the metric (\ref{GeneralMetric}) and in the bosonic contribution to the dilaton (\ref{BosDilaton1}). While these objects have been computed in \cite{SfetsosAdS5}, to evaluate the fermionic contribution to the dilaton we will have to use a different parameterization, so we begin with specifying our coordinates, finding the metric and the bosonic dilaton 
for them, and comparing the results with \cite{SfetsosAdS5}. The fermionic contribution to the dilaton will be evaluated in section \ref{SecFerDilAdS5}.

\subsection{Metric and the bosonic dilaton}
\label{SectAdS5metr}

To apply the procedure outlined in section \ref{SecReview}, we need an explicit form of the coset (\ref{AdS5BosCoset}). The most natural way to parameterize the sphere $S^5=SO(6)/SO(5)$ is to use the Euler angles, and such description has been used in \cite{SfetsosAdS5}, but unfortunately these coordinates make the evaluation of the ferminonic contribution to the dilaton nearly impossible. Thus we use the alternative coordinates introduced in \cite{Bars,SfetsosAdS5}, in which all expressions remain algebraic\footnote{It appears that the authors of \cite{SfetsosAdS5} used the same coordinates while computing the metric and rewrote the final answers in terms of the Euler's angles. We find the algebraic coordinates more convenient.}.

Specifically, we write the element of $SO(6)$ as
\bea\label{SO6gauge}
g=
\left[\begin{array}{cc}
1&0\\
0&{h_m}^n
\end{array}\right]
\left[\begin{array}{cc}
b-1&bX^j\\
-bX_i&\delta_i^j-bX_iX^j
\end{array}\right]\,,\qquad b=\frac{2}{{1+X^m X_m}},
\eea 
where $X^i$ is a five--dimensional vector and ${h_m}^n$ is an element of $SO(5)$. The defining condition for $SO(5)$, $h^T h=I$, can be solved by writing $h$ in terms of an anti-symmetric matrix $A$ as
\bea
{h_m}^n={[(1+A)(1-A)^{-1}]_m}^n\,.
\eea
The $SO(5)$ rotations act on $A$ and $X$ as
\bea
A\rightarrow \Lambda A\Lambda^{-1},\qquad X\rightarrow \Lambda X.
\eea
To fix this gauge freedom, we follow the procedure discussed in \cite{Bars}: first we rotate $A$ to a block form\footnote{Notice that there is a slight difference in gauge fixing between $SO(n)/SO(n-1)$ for odd and even $n$: matrix $A$ has $[(n-1)/2]$ independent components, and there are $[n/2]$ independent $X$.}:
\bea\label{SO6alphaGg}
A=\left(\begin{array}{ccccc}
0&a&0&0&0\\
-a&0&0&0&0\\
0&0&0&b&0\\
0&0&-b&0&0\\
0&0&0&0&0\\
\end{array}\right)\,,
\eea
and then we use the remaining $[SO(2)]^2$ rotations to set $X_2=X_4=0$. 

The $\mathfrak{so}(6)$ algebra has $15$ generators, first ten of them form $\mathfrak{so}(5)$, while the last five correspond to the coset. Specifically, in our parameterization, the coset generators are\footnote{Recall that throughout this article we use hermitian generator, so the element of a group is constructed as $g=\exp[iT x]$.} 
\bea\label{CosetGenerSO6}
(T_\alpha)_{mn}=-\frac{i}{\sqrt{2}}\left[\delta_{m1}\delta_{n(\alpha-9)}-\delta_{n1}\delta_{m(\alpha-9)}
\right]\qquad \alpha=11,\dots 15.
\eea
Application of the procedure (\ref{MainCoset}) leads to the bosonic contribution to the dilaton (\ref{BosDilaton1}) 
\bea\label{AdS5bosDil}
e^{-2\Phi_B}&=&\frac{1024a^2b^2(a^2-b^2)^2X_1^2X_3^2}{(1+a^2)^3(1+b^2)^3 X^2}\, 
\frac{(1-\la^2)^3(1+\la^2)^2}{\la^{10}},
\eea
where we defined
\bea
X^2&\equiv&1+X_1^2+X_3^2+X_5^2.\nonumber
\eea
Note that the lambda dependence factorizes in (\ref{AdS5bosDil}), and this is a general feature of the bosonic dilaton, as discussed in the end of section \ref{SecReview}. Specifically, in the present case, matrix ${\bf P}$ defined in (\ref{TriangM}) has the form
\bea
{\bf P}&=&\left[\begin{array}{ccc}
W(a)&0&0\\
0&W(b)&0\\
0&0&1
 \end{array}\right],\quad \mbox{where}\quad W(x)\equiv (1+x^2)^{-1}\left[\begin{array}{cc}1&-x\\-x& -1 \end{array}\right].
 \eea
This matrix has eigenvalues $\la^{-2}\pm 1$ and
\bea
\mbox{det}\, {\bf P}=\frac{(1-\la^2)^3(1+\la^2)^2}{\la^{10}}\,.
\eea

The metric for $\la$--deformation is constructed using (\ref{MainCoset}), and the result reads
\bea
ds^2_{(\lambda)}&=&\sum_\alpha (e^\alpha_{(\lambda)})^2,\quad 
e^\alpha_{(\lambda)}=\frac{\sqrt{k(1-\la^4)}}{2\la^2}{[{\bf P}^{-1}]^\alpha}_\beta e^\beta_{(0)},
\eea
where $e^\beta_{(0)}$ refer to the frames describing the gauged WZW model ($\la=0$):
\bea\label{WZWframesAdS5}
e_{(0)}^{6}&=&\frac{a^2(1+b^2)X_3^2+(a^2-b^2)X_5^2}{a(1+a^2)(a^2-b^2)X_1}da+\frac{(1+a^2)bX_1}{(a^2-b^2)(1+b^2)}db\nn
&&\qquad+\frac{1}{X^2}\left[-(X^2-X_1^2)dX_1+X_1X_3 dX_3+X_1X_5 dX_5\right],\nn
e_{(0)}^7&=&\frac{da}{X_1(1+a^2)},\quad e_{(0)}^9=\frac{db}{X_3(1+b^2)},\\
e_{(0)}^{10}&=&-\frac{X_5 da}{a(1+a^2)}-\frac{X_5 db}{b(1+b^2)}+\frac{1}{X^2}\left[X_1X_5 dX_1+X_3X_5 dX_3-(X^2-X_5^2)dX_5\right],\nn
e_{(0)}^8&=&-\frac{a(1+b^2)X_3 da}{(1+a^2)(a^2-b^2)}-\frac{(1+a^2)b^2X_1^2-(a^2-b^2)X_5^2}{b(1+b^2)(a^2-b^2)X_3}db\nn
&&\qquad+\frac{1}{X^2}\left[X_1X_3 dX_1-(X^2-X_3^2)dX_3+X_3X_5 dX_5 \right].\nonumber
\eea
The $AdS_5$ counterparts of the metric and the dilaton are obtained by an analytic continuation
\bea\label{AnalContAdS5}
X_1\to i X_1,\quad X_3\to i X_3,\quad k\to -k,
\eea
and the corresponding frames are denoted by $e^1_{(0)}$,\dots,$e^5_{(0)}$.

\subsection{Fermionic dilaton: general discussion}
\label{SecFerDilAdS5}

Although the $SO(6)/SO(5)$ representation (\ref{SO6gauge}) of the five--dimensional sphere is very intuitive, the construction of the supercoset (\ref{AdS5suCoset}) requires embedding of $SO(6)$ into $SU(4)$ and identifying the fermionic degrees of freedom corresponding to the supercoset. We begin with finding the $SU(4)$ matrices in parameterization (\ref{SO6gauge}). 

The $SU(4)$ matrices $g$ describe a representation of $SO(6)$, which acts on anti--symmetric $4\times 4$
matrices $A$ as 
\bea
A\rightarrow gAg^T\,.
\eea 
Specifically, starting with the fundamental representation of $SO(6)$ acting on six--dimensional vectors $(x_1,x_2,x_3,y_1,y_2,y_3)$, one can construct matrix $A$ as
\bea\label{AmatrSU4}
A=\left[\begin{array}{cccc}
0&x_3-iy_3&-x_2+iy_2&x_1+iy_1\\
-x_3+iy_3&0&x_1-iy_1&x_2+iy_2\\
x_2-iy_2&-x_1+iy_1&0&x_3+iy_3\\
-x_1-iy_1&-x_2-iy_2&-x_3-iy_3&0
\end{array}\right]\,.
\eea
The generators of $SU(4)$ are hermitian $4\times 4$ matrices, and to proceed with the coset construction, we need to identify the elements $t_\alpha$ corresponding to the generators (\ref{CosetGenerSO6}). Comparing the action $T_\alpha$ on $(x_1,x_2,x_3,y_1,y_2,y_3)$ and the action of $g\in \mathfrak{su}(4)$  on (\ref{AmatrSU4}), we find
\bea\label{CosetGenerSU4}
T^\alpha c_\alpha=\frac{1}{2}\left[\begin{array}{cccc}
c_{13}&c_{14}+ic_{11}&c_{15}+ic_{12}&0\\
c_{14}-ic_{11}&-c_{13}&0&-c_{15}-ic_{12}\\
c_{15}-ic_{12}&0&-c_{13}&c_{14}+ic_{11}\\
0&ic_{12}-c_{15}&c_{14}-ic_{11}&c_{13}
\end{array}\right]\,.
\eea
All generators of $SU(4)$, including (\ref{CosetGenerSU4}), are hermitian, while generators of $SU(2,2)$ satisfy the modified hermiticity relation
\bea
(T_A)^\dagger=\Sigma T_A\Sigma,\qquad 
\Sigma=\left[\begin{array}{cccc}
0&\sigma_3\\
\sigma_3&0
\end{array}\right].
\eea
For example, the counterparts of the coset generators (\ref{CosetGenerSU4}) are obtained by an analytic continuation
\bea
c_{11}\rightarrow i{\tilde c}_{11},\quad c_{12}\rightarrow i{\tilde c}_{12},\quad c_{13}\rightarrow i{\tilde c}_{13},\quad c_{14}\rightarrow i{\tilde c}_{14},\quad c_{15}\rightarrow {\tilde c}_{15}.
\eea

To proceed we need to construct an automorphism $J_1$ which satisfies (\ref{JprojBos}) for all generators $g\in \mathfrak{su}(4)$ with the exception of (\ref{CosetGenerSU4}). While it is easy to find this $J_1$ for $6\times 6$ matrices and coset generators (\ref{CosetGenerSO6}) (specifically, $J_1=\pm\mbox{diag}(-1,1,1,1,1)$), such matrix does not exist in the four--dimensional representation of $\mathfrak{so}(6)$, and the closest analog of (\ref{JprojBos}) is
\bea\label{JprojBosT}
J_1^{-1} {g}J_1={g^T},\quad 
J_1=\left[\begin{array}{cccc}
0&0&0&1\\
0&0&1&0\\
0&-1&0&0\\
-1&0&0&0
\end{array}\right]\,.
\eea
This means that condition (\ref{SuProj}) will be modified as  
\bea\label{SuProjT}
{\cal P}^{-1}{\cal M}{\cal P}={\cal M}^T,
\eea
and such grading is a familiar feature of $PSU(2,2|4)$ (see, for example, \cite{ArutFrolov} for a detailed discussion). In our parameterization,
\bea\label{PforSU4}
{\cal P}=\left[\begin{array}{cc}
J_1&0\\
0&J_1
\end{array}\right]\,,
\eea
and the detailed discussion of fermions projected out by (\ref{PforSU4}) and relation to other conventions used in the literature is presented in the Appendix \ref{AppPSU}. Here we only mention that if $8\times 8$ supercoset matrix is written as
\bea
{\cal M}=\left[\begin{array}{cc}
A&B\\C&D
\end{array}\right]\,,\quad 
B\equiv\left[\begin{array}{cccc}
b_1&b_2\\
b_3&b_4
\end{array}\right],\quad 
C\equiv\left[\begin{array}{cccc}
c_1&c_2\\
c_3&c_4
\end{array}\right]=-i\left[\begin{array}{cccc}
b^\dagger_3\sigma_3&b^\dagger_1\sigma_3\\
b^\dagger_4\sigma_3&b^\dagger_2\sigma_3
\end{array}\right],
\eea
then projector $P_3$ entering $P_\la$ (\ref{PlaDef}) selects the components satisfying an additional relation (\ref{qqq2}):
\bea
C=\left[\begin{array}{rr}
-[\sigma_1 b_4\sigma_1]^T&[\sigma_1 b_2\sigma_1]^T\\
\ [\sigma_1 b_3\sigma_1]^T&[\sigma_1 b_1\sigma_1]^T
\end{array}\right]\,.
\eea

The last ingredient for constructing the fermionic contribution to the dilaton is the explicit expression for the element of $SU(4)/SO(5)$ in the gauge (\ref{SO6gauge}), (\ref{SO6alphaGg}):
\bea\label{gSAdS5}
g_S&=&\frac{1}{\Delta_S}
\left[\begin{array}{cccc}
1&b&ab&a\\
-b&1&a&-ab\\
a b &-a&1&-b\\
-a &-a b&b&1
\end{array}\right]
\left[\begin{array}{cccc}
1-iX_3&X_1&-iX_5&0\\
-X_1&1+iX_3&0&iX_5\\
-iX_5&0&1+iX_3&X_1\\
0 &iX_5&-X_1&1-iX_3
\end{array}\right]\nn
\phantom{\frac{a}{b}} \\
\Delta_S&=&\sqrt{1+a^2}\sqrt{1+b^2}\sqrt{1+(X_1)^2+(X_3)^2+(X_5)^2}\,.\nonumber
\eea
The element of $SU(2,2)/SO(4,1)$ is obtained by making the analytic continuation (\ref{AnalContAdS5}) in the last expression. Notice that the symmetry
\bea
X_1\leftrightarrow X_3,\quad a\leftrightarrow b,
\eea
which was obvious in the $SO(6)$ parameterization (\ref{SO6gauge}), (\ref{SO6alphaGg}), is less explicit in (\ref{gSAdS5}). 

Evaluation of the fermionic contribution to the dilaton involves a straightforward but tedious calculation of the determinant
\bea
{\mbox{det}[(Ad_f-1-(\lambda^{-2}-1)P_\lambda)|_{\hat{f}_1\oplus \hat{f}_3}]},
\eea
and the results are rather complicated. We collect them and discuss some of their features in the next two subsections. 

\subsection{Dilaton for the gauged WZW model}
\label{SecAdS5WZW}

Geometry with $\la=0$ describes the gauged WZW model, and the solution in this case is given by the frames (\ref{WZWframesAdS5}), along with their AdS$_5$ counterpart and the dilaton
\bea\label{WZWforAdS5}
e^{-2\Phi}&=&2^{20}\frac{a^2b^2(a^2-b^2)^2X_1^2X_3^2}{(1+a^2)^3(1+b^2)^3 X^2}
\frac{{\tilde a}^2{\tilde b}^2({\tilde a}^2-{\tilde b}^2)^2{\tilde X}_1^2{\tilde X}_3^2}{
(1+{\tilde a}^2)^3(1+{\tilde b}^2)^3 {\tilde X}^2}
\left[\frac{X^{2}{\tilde X}^{2}}{X^2+{\tilde X}^2-X^2{\tilde X}^2}\right]^8\\
&&X^2=1+X_1^2+X_3^2+X_5^2,\quad {\tilde X}^2=1-{\tilde X}_1^2-{\tilde X}_3^2+{\tilde X}_5^2\,.\nonumber
\eea
The bosonic contribution to the dilaton is obtained by dropping the expression in the brackets, and the bosonic coset does not require any Ramond--Ramond fluxes. The situation for the supercoset is different, as we have already seen in the 
AdS$_3\times$S$^3$ case: the Ramond--Ramond fluxes are turned on even at $\la=0$. In the present case we were not able to construct the fluxes explicitly, but we verified that the solution (\ref{WZWframesAdS5})--(\ref{WZWforAdS5}) can be supported by $F_5$. 

Recall that the stress--energy tensor for the self--dual five--form,
\bea
T_{mn}=\frac{1}{96}F_{mabcd}{F_n}^{abcd}
\eea
satisfies the Rainich conditions \cite{Rainich}\footnote{For a recent discussion of the original Rainich conditions for electromagnetism and their generalizations to higher dimensions see, for example, \cite{Rainich1,LRT}.} :
\bea\label{Rainich}
{T_m}^m\equiv \mbox{Tr}\, T=0,\quad \mbox{Tr}\, T^3=0,\quad \mbox{Tr}\, T^5=0,\quad \mbox{Tr}\, T^7=0,\quad 
\mbox{Tr}\,  T^9=0,\quad 
\eea
and for geometry supported only by the dilaton and the metric the $T_{mn}$ can be expressed as\footnote{In this paper we are working in the string frame.}
\bea
T_{mn}=R_{mn}+2\nabla_m\nabla_n \Phi.
\eea
The right--hand side vanishes for the ``bosonic'' dilaton, while for the full solution (\ref{WZWforAdS5}) it gives a nontrivial result which satisfies the constraints (\ref{Rainich}). It would be very interesting to find the corresponding flux $F_5$.

\subsection{Special cases for $\la\ne 0$}
\label{SecAdS5SpecCases}

Although the dilaton for arbitrary values of $\la$ can be computed by evaluating the appropriate determinants, unfortunately the results are not very illuminating. In this subsection we will collect the answers for some special cases which give manageable expressions. Since the general expression for the bosonic dilaton is already given by (\ref{AdS5bosDil}), we will focus only on the fermionic contribution to (\ref{SuDilaton1}):
\bea\label{SuDilatonFerm}
e^{2\Phi_F}&=&{\mbox{det}[(Ad_f-1-(\lambda^{-2}-1)P_\lambda)|_{\hat{f}_1\oplus \hat{f}_3}]}\,.
\eea

First we observe that at $\la=0$ the expression for $e^{2\Phi_F}$ depends only on $X_k$ and ${\tilde X}_k$. While this property does not hold for general values of $\la$, setting $a={\tilde a}=b={\tilde b}$ we still find an interesting result:
\bea
e^{2\Phi_F}\Big|_{a={\tilde a}=b={\tilde b}=0}=\left[\frac{(1-\mu X{\tilde X})^2-(1-X^2)(1-{\tilde X}^2)}{X^2{\tilde X}^2}\right]^8,\qquad \mu\equiv \frac{2\la}{\la^2+1}\,.
\eea
In the opposite case, where all $X$ are switched off, the expression is much more complicated, for example at $\la=1$ it has the form
\bea
&&e^{2\Phi_F}\Big|_{X_m={\tilde X}_m=0,\la=1}=F^{12}\left[8F-2+(FP_{-1}-2)^2-2(P_1-1)^2+2P_2\right]^2\,,\\
&&F\equiv(a^2+1)(b^2+1)({\tilde a}^2+1)({\tilde b}^2+1),\nn
&&P_k\equiv (a^2+1)^k+(b^2+1)^k+({\tilde a}^2+1)^k+({\tilde b}^2+1)^k.\nonumber
\eea
In particular, we observe that the last expression is fully symmetric under interchanging the elements of the list $(a,b,{\tilde a},{\tilde b})$. This property persists for all values of $\la$, as long as $X_m={\tilde X}_m=0$, but the general expression is not very illuminating, so we will not write it here. 

The last two interesting cases corresponds to looking only at the sphere or only at the AdS space:
\bea
e^{2\Phi_F}\Big|_S&=&\left[\frac{(AB-\mu X)^2+\mu^2[(AbX_3)^2+(aBX_1)^2-a^2b^2 X^2]}{A^2B^2X^2} \right]^8\,,\nn
e^{2\Phi_F}\Big|_{AdS}&=&\left[\frac{({\tilde A}{\tilde B}-\mu {\tilde X})^2-\mu^2[
({\tilde A}{\tilde b}{\tilde X}_3)^2+({\tilde a}{\tilde B}{\tilde X}_1)^2+{\tilde a}^2{\tilde b}^2 {\tilde X}^2]}{{\tilde A}^2{\tilde B}^2{\tilde X}^2} \right]^8\,,\\
A&=&\sqrt{1+a^2},\quad B=\sqrt{1+b^2},\quad {\tilde A}=\sqrt{1+{\tilde a}^2},\quad {\tilde B}=\sqrt{1+{\tilde b}^2}\,.\nonumber
\eea 
The complexity of our results beyond $\la=0$ suggests that the full solution for the $\la$--deformation of AdS$_5\times$S$^5$ cannot be constructed unless one finds better coordinates, and we leave this problem for future investigation.

\section{Discussion}

In this article we have constructed the supergravity background describing the $\la$--deformation of AdS$_3\times$S$^3$ supercoset and reported some progress towards the analogous result for AdS$_5\times$S$^5$. Our main result is summarized by equations (\ref{AdS3metr}), (\ref{S3metr}),
(\ref{SuDilatonAdS3}), (\ref{C2forAdS3}). In the AdS$_5\times$S$^5$ case we have constructed the metric and the dilaton describing the supercoset, and while the results presented in section \ref{SectAdS5} are rather complicated, there are striking similarities with lower--dimensional cases. For example, at the WZW point, where the expression (\ref{WZWforAdS5}) for the  ten--dimensional dilaton is rather simple, one finds a very close analogy with the six--dimensional case (\ref{QforWZW3}), and we hope that a further exploration of such analogies will lead to construction of full gravity solution for the deformed AdS$_5\times$S$^5$. 

\section*{Acknowledgments}

We thank Arkady Tseytlin and Linus Wulff for useful discussions and for comments on the manuscript. This work is supported in part by NSF grant PHY-1316184.
\appendix

\appendix


\section{$\lambda$--deformation for AdS$_2\times $S$^2$}
\label{AppAdS2}

\renewcommand{\theequation}{A.\arabic{equation}}
\setcounter{equation}{0}

For comparison with the results obtained in this article, we review the geometry of $\lambda$--deformed $AdS_2\times S^2$ constructed in \cite{BTW}. We also extend the solution of \cite{BTW} by one free parameter which makes the fluxes symmetric between the sphere and AdS space.  Applying the procedure reviewed in section \ref{SecReview} to a coset $SU(2)/U(1)$, the authors of \cite{BTW} constructed the metric and the supercoset version of the dilaton (\ref{SuDilaton})\footnote{The solution  corresponding to the bosonic dilaton (\ref{BosDilaton1}) had been constructed earlier in \cite{ST}.}:
\bea\label{AdS2HT}
ds^2&=&\frac{-dx^2+dy^2}{1-\kappa x^2+\kappa^{-1} y^2}+\frac{dp^2+dq^2}{1-\kappa p^2-\kappa^{-1} q^2},\nn
e^{\Phi}&=&\frac{\kappa-x^2+y^2-p^2-q^2+2\sqrt{1-\kappa^2}xp}{\sqrt{-(1-\kappa x^2+\kappa^{-1}y^2)}\sqrt{1-\kappa p^2-\kappa^{-1}q^2}},
\eea
where
\bea
\kappa=\frac{1-\lambda^2}{1+\lambda^2}.
\eea
This background is supported by the Ramond--Ramond flux 
\bea\label{FluxForAdS2}
A&=&\frac{c_1}{M} [ydx-(x-\sqrt{1-\kappa^2}p)dy]+ \frac{c_2}{M}  [q dp-(p-\sqrt{1-\kappa^2}x) dq],\\
M&=&\frac{\kappa-x^2+y^2-p^2-q^2+
2\sqrt{1-\kappa^2}xp}{\sqrt{-(1-\kappa x^2+\kappa^{-1}y^2)(1-\kappa p^2-\kappa^{-1}q^2)}},
\quad c_1^2+c_2^2= 4 \kappa^{-1},\nonumber
\eea
which solves the supergravity equations
\bea
R_{mn}+2\nabla_m\nabla_n \Phi=\frac{e^{2\Phi}}{2}(F_{mp}F_n{}^p-\frac{1}{4}g_{mn}F_{kl}F^{kl}),\nn
\d_n(\sqrt{-g}F^{mn})=0,\quad \nabla^2 e^{-2\Phi}=0,
\eea
and article \cite{BTW} presented the answer (\ref{FluxForAdS2}) for $c_2=0$. 

It is interesting that the flux (\ref{FluxForAdS2}) has a free parameter which interpolates between the components on the sphere and on AdS, while the AdS$_3\times$S$^3$ solution (\ref{C2forAdS3}) has no freedom. This difference can already be seen for the undeformed AdS$_p\times$S$^p$, and it can be traced to the different structure of ``electric--magnetic'' duality groups in four and six dimensions ($U(1)$ in 4d vs $Z_2$ in 6d). 

Since in this article we use parameterization of cosets in terms of $X,A$ coordinates introduced in (\ref{SO6gauge}), we will conclude this appendix by writing the relations between coordinate systems used in \cite{ST,BTW} and a three--dimensional version of (\ref{SO6gauge})--(\ref{SO6alphaGg}) describing $SO(3)/SO(2)$: 
\bea\label{SO3gauge}
g_{so}&=&
\left[\begin{array}{cc}
1&0\\
0&(1+A)(1-A)^{-1}
\end{array}\right]
\left[\begin{array}{ccc}
b-1&bX&0\\
-bX&1-bX^2&-bX\\
0&-bX&1\\
\end{array}\right]\,,\\
A&=&\left[\begin{array}{cc}
0&a\\
-a&0
\end{array}\right],\quad 
b=\frac{2}{{1+X^2}}.\nonumber
\eea 
To compare this with the parameterization in terms of the Euler's angles used in \cite{ST,HT},
\bea
g^{trig}=\exp[i(\phi_1-\phi_2)\sigma_3/2]\exp(i\omega \sigma_2)\exp[i(\phi_1+\phi_2)\sigma_3/2]\,,
\eea
we follow the procedure outlined in section  \ref{SecAltParamAdS3}. Specifically, computing
the matrix $D$ (\ref{DabMap}) and comparing the result with a general parameterization (\ref{SO6gauge}) applied to $SO(3)$, we find 
\bea
X_1&=&-\frac{4(\cos^2\omega\sin2\phi_1+\sin^2\omega\sin2\phi_2)}{4+\cos[2(\omega-\phi_1)]+\cos[2(\omega+\phi_1)]+2\cos2\phi_1+4\cos2\phi_2 \sin^2\omega},\nn
X_2&=&-\frac{4\sin2\omega\sin(\phi_1-\phi_2)}{4+\cos[2(\omega-\phi_1)]+\cos[2(\omega+\phi_1)]+2\cos2\phi_1+4\cos2\phi_2 \sin^2\omega},\nn
a&=&\frac{\cos\phi_2\tan\omega}{\cos\phi_1}.
\eea
A $U(1)$ gauge transformation relates this to (\ref{SO3gauge}) with
\bea
X&=&-\frac{4\sqrt{\sin^2(2\phi_1)+\sin^2(\phi_1-\phi_2)\sin^2(2\omega)}}{4+\cos[2(\omega-\phi_1)]+\cos[2(\omega+\phi_1)]+2\cos2\phi_1+4\cos2\phi_2 \sin^2\omega},\nn
a&=&\frac{\cos\phi_2\tan\omega}{\cos\phi_1}.
\eea
The authors of \cite{ST} fixed the gauge by setting $\phi_2=0$, while the authors of \cite{BTW} chose $\phi_2=\phi_1$ and changed coordinates as 
\bea
\omega=\arccos\sqrt{\kappa p^2+\kappa^{-1}q^2},\qquad \phi_1=\arccos\frac{\sqrt{\kappa}p}{\sqrt{\kappa p^2+\kappa^{-1}q^2}}
\eea
to arrive at (\ref{AdS2HT}).


\section{Parametrization of $\mathfrak{psu}(1,1|2)$ and $\mathfrak{psu}(2,2|4)$}
\label{AppPSU}

\renewcommand{\theequation}{B.\arabic{equation}}
\setcounter{equation}{0}

In this appendix we briefly summarize the parameterization of $\mathfrak{psu}(1,1|2)$, $\mathfrak{psu}(2,2|4)$, and their cosets used in sections \ref{SecAdS3} and \ref{SectAdS5}. We will mostly follow the notation of \cite{ArutFrolov,Beisert}, although our parameterization of fermions differs from the one in \cite{ArutFrolov}, and we will comment on the difference.

The Lie superalgebras $\mathfrak{psu}(n,n|2n)$ can be defined in terms of $(4n)\times (4n)$ supermatrices
\bea\label{GenSuAlg}
{\cal M}=\left[\begin{array}{cc}
A&B\\C&D
\end{array}\right]\,,
\eea
with even $(2n)\times (2n)$ blocks $A$, $D$ and odd $(2n)\times (2n)$ blocks $B$, $C$. The graded Lie bracket is defined as
\bea
[{\cal M},{\cal M}'\}=\left[\begin{array}{cc}
AA'+BC'-A'A+B'C&AB'+BD'-A'B-B'D\\CA'+DC'-C'A-D'C&CB'+DD'+C'B-D'D
\end{array}\right]\,,
\eea
Matrix ${\cal M}$ is subject to the hermiticity condition
\bea\label{HermitPSU}
\left[\begin{array}{cc}
A&B\\C&D
\end{array}\right]=\left[\begin{array}{cc}
\Sigma A^\dagger \Sigma^{-1}&-i\Sigma C^\dagger\\-iB^\dagger \Sigma^{-1}&D^\dagger
\end{array}\right]\,,
\eea
where $\Sigma$ is a hermitian matrix of signature $(n,n)$. Convention for $\mathfrak{su}(n,n)$ represented by $A$ fixes the matrix $\Sigma$ and the parameterization of fermions $B$, $C$.

\bigskip

For $\mathfrak{psu}(1,1|2)$ we choose $\Sigma=\mbox{diag}(1,-1)$. This leads to the relation
\bea
C=-iB^\dagger\left[\begin{array}{rr}
1&0\\
0&-1
\end{array}\right],
\eea
or more explicitly
\bea\label{BCcomponents}
B=\left[\begin{array}{rr}
b_{11}&b_{12}\\
b_{21}&b_{22}
\end{array}\right],\quad C=\left[\begin{array}{rr}
-ib^\dagger_{11}&ib^\dagger_{21}\\
-ib^\dagger_{12}&ib^\dagger_{22}
\end{array}\right]\,.
\eea
To construct the algebra for the coset 
\bea
\frac{PSU(1,1|2)_l\times PSU(1,1|2)_r}{SU(1,1)_{diag}\times SU(2)_{diag}},
\eea
we take two copies of $\mathfrak{psu}(1,1|2)$,
\bea\label{CalMwrngBlcks}
{\cal M}'=\left[\begin{array}{cc}
{\cal M}_1&0\\
0&{\cal M}_2
\end{array}\right]\,,
\eea
and project to the subgroup $H$ by imposing the relation (\ref{SuProj})
\bea\label{SuProjApp}
{\cal P}^{-1}{\cal M}'{\cal P}={\cal M}'\,,
\eea
as discussed in section \ref{SecAdS3ferm}. Notice that AdS$_3$ and S$_3$ blocks are mixed in the matrix (\ref{CalMwrngBlcks}), and to make the separation more explicit we rearrange the components of the matrix ${\cal M}'$ using the parameterization (\ref{GenSuAlg}) for ${\cal M}_1$ and ${\cal M}_2$. Specifically we define
\bea\label{MAdS3Split}
{\cal M}=\left[\begin{array}{cc|cc}
A_1&0&B_1&0\\
0&A_2&0&B_2\\
\hline
C_1&0&D_1&0\\
0&C_2&0&D_2
\end{array}\right]\,.
\eea
The top left block of this matrix describes AdS space, the bottom right block describes the sphere, and the matrix ${\cal P}$ corresponding to this supercoset is given by (\ref{CalPforAdS3}):
\bea\label{AdS5Papp}
{\cal P}=\left[\begin{array}{cccc}
0&1_{2\times 2}&0&0\\ 
1_{2\times 2}&0&0&0\\
0&0&0&1_{2\times 2}\\ 
0&0&1_{2\times 2}&0
\end{array}\right]\,.
\eea
In particular, this matrix does not mix the $B_i$ and $C_i$ components, so in section \ref{SecAdS3ferm} we computed the fermionic contribution to the dilaton by treating the holomorphic and anti--holomorphic components ($b_{ij}$ and $b^\dagger_{ij}$) as independent variables.

\bigskip

Let us now discuss the $\mathfrak{psu}(2,2|4)$ superalgebra, which emerges in the description of strings on AdS$_5\times$S$^5$ \cite{AdS5PSU}. In this case equation (\ref{HermitPSU}) involves $4\times 4$ blocks, and we choose the matrix $\Sigma$ involved in the hermiticity condition (\ref{HermitPSU}) to be 
\bea
\Sigma=\left[\begin{array}{cccc}
0&\sigma_3\\
\sigma_3&0
\end{array}\right]
\eea
This choice leads to a relation between $2\times 2$ blocks of $B$ and $C$ in (\ref{GenSuAlg}):
\bea\label{qqq0}
B\equiv\left[\begin{array}{cccc}
b_1&b_2\\
b_3&b_4
\end{array}\right],\quad 
C\equiv\left[\begin{array}{cccc}
c_1&c_2\\
c_3&c_4
\end{array}\right]=-i\left[\begin{array}{cccc}
b^\dagger_3\sigma_3&b^\dagger_1\sigma_3\\
b^\dagger_4\sigma_3&b^\dagger_2\sigma_3
\end{array}\right],
\eea
A choice of holomporhic and anti--holomorphic fermions is no longer convenient since the coset projection mixes them. As discussed in section \ref{SecFerDilAdS5}, for the $\mathfrak{psu}(2,2|4)$ supercoset, the condition (\ref{SuProj}) is replaced by  (\ref{SuProjT})
\bea\label{qqq1}
{\cal P}^{-1}{\cal M}{\cal P}={\cal M}^T,
\eea
with ${\cal P}$ given by (\ref{PforSU4}). An explicit calculation shows that projection (\ref{qqq1}) chooses the elements which satisfy 
\bea\label{qqq2}
B=\left[\begin{array}{cccc}
b_1&b_2\\
b_3&b_4
\end{array}\right],\quad 
C=\left[\begin{array}{rr}
-[\sigma_1 b_4\sigma_1]^T&[\sigma_1 b_2\sigma_1]^T\\
\ [\sigma_1 b_3\sigma_1]^T&[\sigma_1 b_1\sigma_1]^T
\end{array}\right]
\eea
in addition to (\ref{qqq0}). The coset corresponds to the generators included in (\ref{qqq0}), but not in (\ref{qqq2}). In other words, generators satisfying both (\ref{qqq2}) and (\ref{qqq0}) survive under projection $P_3$, and $P_1$ is defined as $P_1=1-P_3$. 

We conclude this appendix by relating our conventions with notation used in \cite{ArutFrolov}. We chose a different embedding of the coset into $SU(4)\times SU(2,2)$, and this led to a following relation between our generators and the ones used by Arutyunov and Frolov (AF) \cite{ArutFrolov}:
\bea
T_{su(4)}&=&R T_{su(4)}^{AF} R^{-1},\quad 
R=\left[\begin{array}{cccc} 
i&0&0&0\\
0&0&0&i\\
0&0&1&0\\
0&1&0&0
\end{array}\right],\\
T_{su(2,2)}&=&\tilde{R} T^{AF}_{su(2,2)} \tilde{R}^{-1},\quad 
\tilde{R}=\frac{1}{\sqrt{2}}\left[\begin{array}{cccc} 
1&0&-1&0\\
0&1&0&1\\
1&0&1&0\\
0&-1&0&1
\end{array}\right].
\eea
While our generators are convenient for evaluating the $\la$--deformation, the generators of Arutyunov and Frolov are better suited for imposing kappa symmetry. Specifically, elimination of this freedom in the notation of  \cite{ArutFrolov} gives
\bea
B^{AF}=\left[\begin{array}{cccc}
0&b_2\\
b_3&0
\end{array}\right],\quad 
C^{AF}=\left[\begin{array}{rr}
0&c_2\\
c_3&0
\end{array}\right]
\eea
while in our notation
\bea
B=\left[\begin{array}{cccc}
b_1&b_2\\
\sigma_3b_1\sigma_3&-\sigma_3 b_2\sigma_3
\end{array}\right],\quad 
C=\left[\begin{array}{rr}
i\sigma_3 b_1^\dagger&ib_1^\dagger\sigma_3\\
-i\sigma_3b^\dagger_2&ib_2^\dagger\sigma_3
\end{array}\right]\,.
\eea
The expressions for kappa symmetry are not used in this paper.

\end{document}